\documentclass[12pt]{article}

\usepackage{graphicx}
\usepackage{amsfonts}

\newcommand\fiverm{\tiny\rm}
\input  ./prepictex
\input  ./pictex
\input  ./postpictex

\begin{document}
\baselineskip1.8em

\setcounter{page}{1}

\begin{center}
\textbf{\large\bf Efficient and Accurate  Calibration to  FX Market Skew\\[3mm]
with Fully Parameterized Local Volatility Model}
        \\[0.8cm] Dongli Wu, Bufan Zhang, Xiao Lin\footnote{E-mail: xiao\_lin\_99@qq.com; First version: 2022-11-15; This version: 2023-04-21.} \\[3mm]
       {\small  CCB Fintech \\[1mm]
         Shanghai 200120, China \\
 		 }
\vspace{4mm}
\end{center}

\vspace{4mm}

{\begin{center}\begin{minipage}{13.9cm}
\baselineskip1.6em
{\sc Abstract}:
When trading American and Asian options in the FX derivatives market, banks must calculate prices using a complex mathematical model. It is often observed that different models produce varying prices for the same exotic option, which violates the non-arbitrage requirement of derivative risk management. To address this issue, we have studied a fully parameterized local volatility model for pricing American/Asian options. This model, when implemented using a grid or Monte-Carlo numerical method, can be efficiently and accurately calibrated to FX market skew volatilities. As a result, the model can provide reliable prices for exotic options during daily trading activities.

\vspace{5mm}


\end{minipage}\end{center}}

\vspace{5mm}

\section{\large Introduction}
\setcounter{equation}{0}

Recently, the China Foreign Exchange Trade System (CFETS) opened two types of FX exotic derivatives business in the interbank market: American options and Asian options. Unlike European options, which are traded by prices quoted in the liquid market, banks must issue prices for exotic options by calculating them using complicated mathematical models. A review paper by Homescu~[1] provides a list of such models. It is a common requirement for banks and financial institutions that a mathematical model for pricing FX derivatives should be arbitrage-free in a risk-neutral market. However, in daily banking practice, it is often observed that different models produce different prices even for a specific exotic option.

To eliminate arbitrage opportunities, it is common practice in the financial industry to calibrate the model before pricing FX derivatives.
The formulation for such calibration can be found in the paper by Derman and Kani~[2]. Calibration is often the most difficult task in model development, particularly when the model has a complex mathematical structure and many free parameters. Calibration is also essential for model validation, as the model's performance during calibration provides a unique standard for judging an arbitrage-free model. Unfortunately, despite a vast number of existing publications, such as [1-7], it is difficult to find a model that efficiently and accurately calibrates to the FX market skew volatility surface.

The purpose of this paper is to study a fully parameterized local volatility model (LV) for efficient and accurate volatility calibration. This LV model delivers satisfactory pricing for our trading desk.

In Section 2, we briefly describe the deal structure for FX American and FX Asian options. The values of these options depend on the evolution of FX market prices throughout the entire life of the deals. Therefore, a model that takes into account the evolution process of FX prices and volatilities is necessary for pricing.

Section 3 presents the mathematical formulation of the fully parameterized local volatility model. The stochastic differential equation is similar to that of a log-normal model, but the volatility is a function of time and the underlying FX price. When performing numerical calculations, the volatility function can be discretized using a set of two-dimensional parameters. It is important to choose a sufficiently large number of parameters to accurately calibrate to the market.

In Section 4, we present a grid numerical method for solving the stochastic LV model equation. This method is most efficient for pricing American options, where the deal value is independent of the historical path of the market FX price.

In Section 5, we present a Monte-Carlo numerical method for solving the stochastic LV model equation. This method is most suitable for pricing Asian options, where the deal value strongly depends on the historical path of the market FX price.

In Section 6, we focus on model calibration. We formulate a mathematical procedure using the Levenberg-Marquardt method to calibrate the LV model parameters. With an example, we show that this procedure can efficiently and accurately calibrate the model parameters for the full FX market volatility matrix. We also provide two conditions for the input market volatility data necessary for successful calibration.

In Section 7, we present two calculation examples to demonstrate the pricing accuracy of American and Asian options using the LV model. We show that the LV model in this paper produces prices that quickly converge to stability when increasing the number of grids or Monte-Carlo paths, and that errors are stable within 1 basis point. This satisfies trading requirements.

Finally, in Section 8, we compare our one-factor LV model to the two-factor LSV (Local-Stochastic Volatility) model, which is currently widely used in the financial industry. Since the LV model in this paper performs exceptionally well in both model calibration and pricing, we suggest that there is no need to extend this model to the two-factor LSV model.

\section{Deal structure of American/Asiain options}

First, we  briefly review the deal structure of two types of exotic options.

The American option is an extension of the European option. In the traditional European option, the option buyer has the right (but not the liability) to buy or sell a certain amount of foreign currency $N$ at a previously determined unit price $K$ on the expiration date $T_e$. In the corresponding American option, the exercise date is not restricted to $T_e$ only, but falls within a time period. For example, in the USDCNY option market, a bank buys an American call option today, $t_0 = 20221012$. The bank has one opportunity to buy USD 1000000 using the price $K=7.1234$ during the time period from $T_s = 20221112$ to $T_e = 20231012$. Let $S(t)$ be the FX price of USDCNY at time $t$. When the option is exercised, the immediate profit in CNY to the bank is
\begin{equation}
            {\rm Payoff}(t) = N \max \big( S(t) - K,\, 0\big).
\end{equation}
The American option requires a human action to be exercised. If the bank does not exercise the option during the time period $T_s \le t \le T_e$, its value becomes zero.

An Asian option has a common expiration date $T_e$ like a European option. However, their payoffs are calculated differently. Let $T_s \le t^1 < t^2 < \cdots < t^n \le T_e$ be a time series in the option lifetime. The time series can be set up in several ways, for example, daily, weekly, or monthly, but all $t^k$ must be a CFETS working date so that there is always an FX rate $S^k = S(t^k),\, (k = 1,2,\cdots,n)$ available for calculating the payoffs. Let
\begin{eqnarray} \label{fixing}
    \bar{S} &=& \frac{1}{n} \big(S^1 + S^2 + \cdots + S^n \big) ,
    \nonumber\\[2mm]
    \hat{S} &=& \Big( S^1 S^2\cdots S^n \Big)^{\frac{1}{n}} .
\end{eqnarray}
There are four types of payoffs in Asian options. Taking the call option as an example, the first is called Arithmetic Spot (all in CNY)
\begin{equation} \label{payoff_asia_Arithmetic_Spot}
    {\rm Payoff}(T_e) = N \max\big( \bar{S} - K, \, 0\big).
    \end{equation}
The second is called Geometric Spot
\begin{equation}
    {\rm Payoff}(T_e) = N \max\big( \hat{S} - K, \, 0\big).
\end{equation}
The third is called Arithmetic Strike
\begin{equation}
    {\rm Payoff}(T_e) = N \max\big(S(T_e) - \bar{S}, \, 0\big).
\end{equation}
The last is called Geometric Strike
    \begin{equation} \label{payoff_asia_Geometric_Strike}
{\rm Payoff}(T_e) = N \max\big( S(T_e) - \hat{S} , \, 0\big).
\end{equation}
In general, if $t^n < T_e$, there may be $S(T_e) \neq S_n$.

\section{Fully parameterized local volatility model}

As the payoffs of American or Asian options depend on the evolution process of the FX rate $S(t)$ over a future time period $T_s \le t \le T_e$, we require a stochastic model to formulate the changes of $S(t)$.

Assuming that $S(t)$ in $0 < t \le T_e$ is governed by a local volatility process:
\begin{equation} \label{SDE}
\frac{{\rm d} S}{S} = \mu\, {\rm d}t + \sigma(t,S)\, {\rm d} w,
\end{equation}
Here, $\mu = \mu(t)$ is the deterministic drift term, $w=w(t)$ is a Brownian motion, and $\sigma(t,S)$ is the volatility (also called model volatility to distinguish it from market volatility in the following). This model is called a local volatility model because $\sigma(t,S)$ is a function of time $t$ and the local FX state $S=S(t)$.
From equation (\ref{SDE}), as long as the function $\sigma(t,S)$ is known, we can solve the stochastic process for $S(t)$.






\renewcommand\arraystretch{1.05}
\begin{table}[t]
\begin{center}
\caption{Fully parameterized local volatility function $\sigma(t,s)/100$.}
\vspace{4mm}
\begin{small}
\begin{tabular}{|c||c|c|c|c|c|c|c|c|c|} \hline
\rule[0pt]{0pt}{5mm}
\tt	  day/s   &\tt 	0.0	    &\tt	 0.1    &\tt  0.2 	&\tt  0.3	&\tt  0.4   &\tt  0.5  &\tt  0.6 &\tt \ $\cdots$\ \ &\tt  1.0 \\
\hline \hline
\tt  0.0	&\tt	4.300	&\tt	4.300	&\tt 4.300	&\tt 4.300	&\tt 4.300	&\tt 7.746 &\tt 4.300 &\tt $\cdots$ &\tt 4.300 \\
\tt	 1.0	&\tt	4.303	&\tt	4.376	&\tt 3.950	&\tt 3.190	&\tt 2.449	&\tt 5.150 &\tt 2.449 &\tt $\cdots$ &\tt 4.317 \\
\tt	 2.5	&\tt	4.299	&\tt	4.292	&\tt 4.322	&\tt 3.947	&\tt 2.630	&\tt 0.828 &\tt 2.630 &\tt $\cdots$ &\tt 4.301 \\
\tt	 4.0	&\tt	4.299	&\tt	4.207	&\tt 3.842	&\tt 3.535	&\tt 2.353	&\tt 0.218 &\tt 2.353 &\tt $\cdots$ &\tt 4.315 \\
\tt	 9.0	&\tt	4.309	&\tt	4.308	&\tt 3.726	&\tt 2.681	&\tt 3.266	&\tt 3.054 &\tt 3.266 &\tt $\cdots$ &\tt 4.336 \\
\tt	14.0	&\tt	4.315	&\tt	4.528	&\tt 4.234	&\tt 3.244	&\tt 3.833	&\tt 3.374 &\tt 3.833 &\tt $\cdots$ &\tt 4.363 \\
\tt	17.5	&\tt	4.309	&\tt	4.515	&\tt 4.885	&\tt 4.396	&\tt 3.679	&\tt 3.775 &\tt 3.679 &\tt $\cdots$ &\tt 4.343 \\
\tt	21.0	&\tt	4.309	&\tt	4.741	&\tt 5.978	&\tt 6.007	&\tt 4.991	&\tt 5.540 &\tt 4.991 &\tt $\cdots$ &\tt 4.356 \\
\tt	25.5	&\tt	4.301	&\tt	4.610	&\tt 6.211	&\tt 6.686	&\tt 6.227	&\tt 5.987 &\tt 6.227 &\tt $\cdots$ &\tt 4.318 \\
\tt	30.0	&\tt	4.310	&\tt	4.344	&\tt 4.954	&\tt 3.995	&\tt 2.646	&\tt 2.803 &\tt 2.646 &\tt $\cdots$ &\tt 4.334 \\
\tt	59.0	&\tt	4.324	&\tt	4.892	&\tt 5.252	&\tt 4.943	&\tt 4.983	&\tt 5.132 &\tt 5.382 &\tt $\cdots$ &\tt 4.455 \\
\tt	91.0	&\tt	4.282	&\tt	4.324	&\tt 5.675	&\tt 5.333	&\tt 4.329	&\tt 4.422 &\tt 4.098 &\tt $\cdots$ &\tt 4.338 \\
\tt	179.0	&\tt	4.290	&\tt	4.204	&\tt 3.783	&\tt 3.902	&\tt 3.763	&\tt 3.124 &\tt 4.156 &\tt $\cdots$ &\tt 4.552 \\
\tt	273.0	&\tt	4.291	&\tt	4.121	&\tt 4.214	&\tt 4.761	&\tt 4.628	&\tt 3.896 &\tt 4.098 &\tt $\cdots$ &\tt 4.462 \\
\tt	365.0	&\tt	4.293	&\tt	4.240	&\tt 3.823	&\tt 4.021	&\tt 4.199	&\tt 3.393 &\tt 4.214 &\tt $\cdots$ &\tt 4.524 \\
\tt	547.0	&\tt	4.293	&\tt	4.167	&\tt 4.125	&\tt 4.854	&\tt 4.751	&\tt 3.773 &\tt 4.336 &\tt $\cdots$ &\tt 4.446 \\
\tt	730.0	&\tt	4.298	&\tt	4.233	&\tt 3.629	&\tt 3.830	&\tt 3.911	&\tt 3.255 &\tt 4.485 &\tt $\cdots$ &\tt 4.469 \\
\tt	1094.0	&\tt	4.298	&\tt	4.254	&\tt 3.976	&\tt 5.202	&\tt 5.454	&\tt 3.303 &\tt 4.859 &\tt $\cdots$ &\tt 4.362 \\
\hline
\end{tabular}
\end{small}
\end{center}
\end{table}


We can discretize the continuous function $\sigma(t,S)$ by a set of two-dimensional parameters. First, we perform a variable transformation for $S$ as follows:
\begin{equation}
        s = \Phi\Big( \frac{ \ln(S/F_t) }{ 1.3c_0 \sqrt{t + 1/365.25} } \Big),
\end{equation}
where $F_t$ is the forward FX value at time $t$ (which can be obtained from the FX curve), $c_0$ is a constant that can be set to the 1-year ATM market volatility, and $\Phi(\cdot)$ is the cumulative probability function of the normal distribution:
\begin{equation}
            \Phi(x) = \frac{1}{\sqrt{2\pi}} \int\limits_{-\infty}^x e^{-\frac{1}{2}y^2}  {\rm d} y.
\end{equation}
Using the variable $s$, we can restrict the definition domain of $\sigma(t,S)$ to a rectangular region:
\begin{equation}
0 \le t \le T_e,  \hspace{1cm}
0 \le s \le 1.
\end{equation}
This region can be divided into a set of small rectangular elements, in which the function is calculated using bilinear interpolation. As an example, Table~1 shows a discrete volatility surface $\sigma(t,s)$ that has been decomposed into $18\times 11$ grids. The number of grids can be increased if the computer's memory and speed improve.

\section{Grid method for American options}

The value of an American option at time $t=T$ is independent of the historical path of the underlying $S(t)$ in $t<T$. Therefore, a grid method can be used to evaluate the American option. To price thousands of deals for the desk, we need to build a symmetric, stable, and flexible grid. In the grid method, we will first ignore the drift term. The work is carried out based on the following model equation:
\begin{equation}
{\rm d} x = \sigma (t,S) {\rm d} w, \hspace{1cm} S = e^{x + \mu},
\end{equation}
where $\mu=\mu(t)$ is a parameter for curve calibration to the FX forward curve.

If a correct volatility function $\sigma(t,S)$ is already known, the entire pricing process in the grid method can be divided into three stages: grid setup, curve calibration by forward propagation, and deal valuation by backward propagation. However, if the correct volatility function $\sigma(t,S)$ is not yet available, we assume a volatility function and use the three-stage process repeatedly to calibrate the volatility until we have the correct volatility function. After that, we can perform the valuation work. In this section, we focus on the three-stage process by assuming that the correct volatility function is available. The volatility calibration is discussed in Section~6.

The grid setup in the first stage includes both a time grid and a state grid.

We begin by setting up the time grid, which divides the time period $t$ $(0 < t < T_e)$ into several small intervals. The process starts by collecting special dates (e.g., expiration dates) from market volatility instruments and American option deals to be priced. These dates form the initial time series. If an interval in the initial time series is too large (for example, greater than 3 days), we add more dates in between to create a finer-grained grid. This completes the time grid setup.





\begin{figure}[hbt]
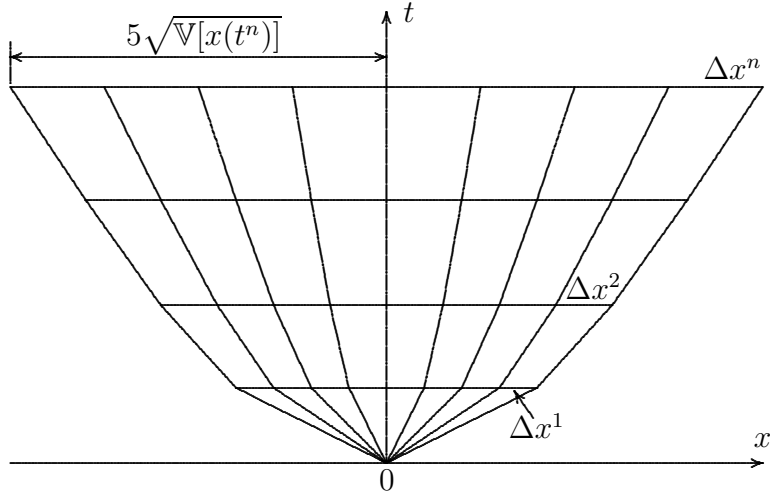


\[  \beginpicture   \setlinear

\setcoordinatesystem units <10mm,10mm>
\setplotarea x from  -5.5 to 5.5, y from 0 to 6

\thinlines

\arrow <1,5mm>   [0.25,0.75] from  -5.0 0 to 5.0  0
\arrow <1,5mm>   [0.25,0.75] from  0.0 0.0 to  0.0 6.0

\plot  -2  1    2  1  /
\plot  -3  2.1  3  2.1  /
\plot  -4  3.5  4  3.5  /
\plot  -5  5  5  5  /

\plot 0 0   -2 1  -3  2.1     -4  3.5    -5  5 /
\plot 0 0    2 1   3  2.1      4  3.5     5  5 /

\plot 0 0  0.5 1   0.75  2.1   1 3.5  1.25 5 /
\plot 0 0  1.0 1   1.5   2.1   2 3.5  2.5 5 /
\plot 0 0  1.5 1   2.25  2.1   3 3.5  3.75 5 /

\plot 0 0  -0.5 1   -0.75  2.1   -1 3.5  -1.25 5 /
\plot 0 0  -1.0 1   -1.5   2.1   -2 3.5  -2.5 5 /
\plot 0 0  -1.5 1   -2.25  2.1   -3 3.5  -3.75 5 /

\put {0} [c] at 0 -0.22
\put {$x$} [c] at 5.0 0.3
\put {$\Delta x^1$} [c] at 2 0.5
\arrow <1,5mm>   [0.25,0.75] from 1.95 0.6 to 1.7 0.95

\put {$\Delta x^2$} [c] at 2.75 2.35
\put {$\Delta x^n$} [c] at 4.6 5.25

\put {$t$} [c] at 0.3 6.0

\plot -5 5.05  -5 5.6 /
\arrow <1,5mm>   [0.25,0.75] from -2 5.4 to -5 5.4
\arrow <1,5mm>   [0.25,0.75] from -2 5.4 to 0 5.4

\put { 5$\sqrt{{\mathbb V} \big[ x(t^n) \big]}$} [c] at -2.5 5.7

\endpicture     \]
\caption{A tree-like grid field.}

\end{figure}


Then we set up the state grid. We denote the time series as $t^0=0, t^1, t^2, \cdots$. We  build a tree-like grid field, similar to the one shown in Figure~1. The half-space width at date $t$ is set to 5 times the standard deviation of $x(t)$, which can be calculated using the variance in the following formula,
\begin{equation}
            {\mathbb V} \big[ x(t^n) \big] = {\mathbb V}\big[ x(t^{n-1}) \big] + \sigma_{\max}^2 \Delta t,
\end{equation}
where, $\sigma_{\max} = \max\limits_{t=t^n} \{ \sigma(t,S(t)) \}$. The half space width is divided into a number of $I$ intervals $\Delta x^n$ (for example, $I \ge 50$). Then, there are  a total of $2I+1$ state grid points at time $t^n$,
\begin{equation}
            x^n_i =5\sqrt{ {\mathbb V} \big[ x(t^n) \big]  }\, \frac{\,i\,}{I} , \hspace{1cm}
                    (i = 0, \pm 1, \pm 2, \cdots, \pm I).
\end{equation}
Each grid point $x_i^n$ is mapped to an FX state, namely
\begin{equation} \label{state_map}
                S^n_i = \exp( x^n_i +\mu^n).
\end{equation}
  

 


\begin{figure}[hbt]
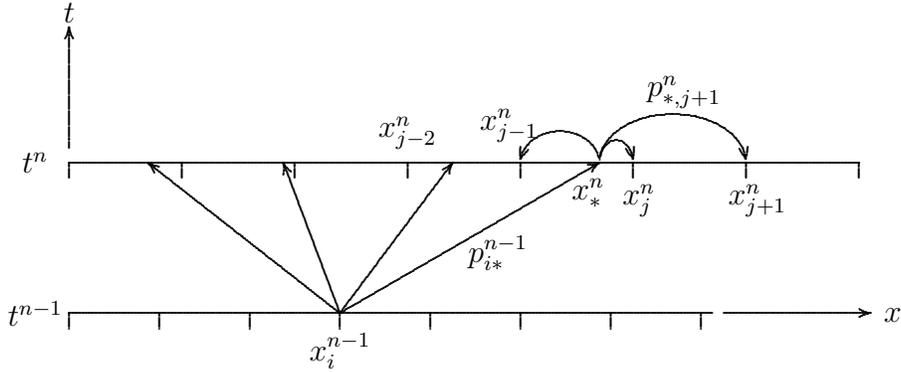


\[ \beginpicture    \setlinear

\setcoordinatesystem units <15mm,20mm>
\setplotarea x from -0.5 to 7.5, y from -0.2 to 1

\plot 0.0  0  5.7 0 / 
\arrow <1,5mm>   [0.25,0.75] from  5.8  0  to 7.1  0
\put {$x$} at 7.3 -0.01

\plot 0.0  0  0.0  -0.1 /
\plot 0.8  0  0.8  -0.1 /
\plot 1.6  0  1.6  -0.1 /
\plot 2.4  0  2.4  -0.1 /
\plot 3.2  0  3.2  -0.1 /
\plot 4.0  0  4.0  -0.1 /
\plot 4.8  0  4.8  -0.1 /
\plot 5.6  0  5.6  -0.1 /

\plot 0 1  7 1 /
\plot 0  1  0  0.9 /
\plot 1  1  1  0.9 /
\plot 2  1  2  0.9 /
\plot 3  1  3  0.9 /
\plot 4  1  4  0.9 /
\plot 5  1  5  0.9 /
\plot 6  1  6  0.9 /
\plot 7  1  7  0.9 /

\arrow <1,5mm>   [0.25,0.75] from  2.4  0  to 3.4 1
\arrow <1,5mm>   [0.25,0.75] from  2.4  0  to 4.7 1
\arrow <1,5mm>   [0.25,0.75] from  2.4  0  to 1.9 1
\arrow <1,5mm>   [0.25,0.75] from  2.4  0  to 0.7 1

\put {$x^{n-1}_i$}    at 2.4 -0.25
\put {$p^{n-1}_{i*}$} at 3.8  0.4
\put {$x^{n}_{*}$}    at 4.6  0.8

\put {$x^{n}_{j}$}    at 5.05  0.75
\put {$x^{n}_{j+1}$}  at 6.1   0.75
\put {$x^{n}_{j-1}$}  at 3.90   1.25
\put {$x^{n}_{j-2}$}  at 3.0    1.2

\plot 
4.9927 	1.0464 
4.9837 	1.0681 
4.9714 	1.0882 
4.9561 	1.1061 
4.9382 	1.1214 
4.9181 	1.1337 
4.8964 	1.1427 
4.8735 	1.1482 
4.8500 	1.1500 
4.8265 	1.1482 
4.8036 	1.1427 
4.7819 	1.1337 
4.7618 	1.1214 
4.7439 	1.1061 
4.7286 	1.0882 
4.7163 	1.0681 
4.7073 	1.0464 
4.7018 	1.0235   /
\arrow <1,5mm>   [0.45,0.75] from 4.9927 	1.0464  to 4.999 1.03

\plot
5.9920 	1.0508 
5.9682 	1.1004 
5.9292 	1.1475 
5.8759 	1.1910 
5.8096 	1.2298 
5.7321 	1.2629 
5.6451 	1.2896 
5.5509 	1.3091 
5.4517 	1.3210 
5.3500 	1.3250 
5.2483 	1.3210 
5.1491 	1.3091 
5.0549 	1.2896 
4.9679 	1.2629 
4.8904 	1.2298 
4.8241 	1.1910 
4.7708 	1.1475 
4.7318 	1.1004 
4.7080 	1.0508   /
\arrow <1,5mm>   [0.45,0.75] from 5.9960 1.04  to 5.999 1.03
\put {$p^{n}_{*,j+1}$} at 5.45 1.47

\plot
4.0043 	1.0329 
4.0171 	1.0649 
4.0381 	1.0953 
4.0668 	1.1234 
4.1025 	1.1485 
4.1443 	1.1699 
4.1911 	1.1871 
4.2418 	1.1997 
4.2952 	1.2074 
4.3500 	1.2100 
4.4048 	1.2074 
4.4582 	1.1997 
4.5089 	1.1871 
4.5557 	1.1699 
4.5975 	1.1485 
4.6332 	1.1234 
4.6619 	1.0953 
4.6829 	1.0649 
4.6957 	1.0329 /
\arrow <1,5mm>   [0.45,0.75] from 4.005 1.04  to 4.0 1.03

\arrow <1,5mm>   [0.25,0.75] from  0 1.1  to 0 1.9
\put {$t$} at 0 2.0

\put {$t^n$} at -0.3 1.0
\put {$t^{n-1}$} at -0.3 0.0

\endpicture     \]
\caption{ Transition probability between grid points.}

\end{figure}


We still need to determine the transition probability in the grid setup. We use $p^{nl}_{ij}$ to denote the transition probability from point $x^n_i$ to $x^l_j$. In case that $l = n+1$, we will omit the $l$ and simply write it as $p^{n}_{ij}$. Then, the transition probability from grid point $x_i^{n-1}$ to  $x_j^n$, namely $p^{n-1}_{ij}$, is calculated in two steps, as shown in Figure~2.

In the first step, a quaternary tree is created to the starting point $x_i^{n-1}$ by
\begin{eqnarray}  \label{4.4}
	\Delta w &=& \Big( -\sqrt{3\Delta t},\, - \sqrt{\frac{\Delta t}{3}},\, \sqrt{\frac{\Delta t}{3}}, \,\sqrt{3\Delta t}\ \Big),
		\nonumber \\[2mm]
	p^{n-1}_{i*} &=&  \Big(\frac{1}{8},\ \frac{3}{8}, \ \frac{3}{8}, \ \frac{1}{8}\ \Big).
\end{eqnarray}
Each tree branch points to one of the four intermediate points $x^n_*$ at time $t^n$, namely
\begin{equation} \label{4.3}
		x^n_* = x^{n-1}_i + \sigma(t^{n-1}, S^{n-1}_i) \cdot \Delta w.
\end{equation}

In the second step, for each intermediate point $x^n_*$, we find a grid point $x^n_j$ that is closest to it. From a Taylor expansion, we can calculate a state variable $S^n_*$ at point $x^n_*$ by combining three neighboring values, as follows:
\begin{equation} \label{4.5}
		S^n_* = \frac{1}{2}\big(\xi^2-\xi \big)\,S^n_{j-1} + \big(1-\xi^2\big)\, S^n_j +
			\frac{1}{2}\big(\xi^2+\xi \big)\,S^n_{j+1},
\end{equation}
where $\xi = (x^n_* - x^n_j)/\Delta x^n$. The coefficients in the formula can be referred to as the transition probabilities from $x^n_*$ to the three neighboring grid points, namely,
\begin{equation} \label{4.6}
		p^{nn}_{*,\,j-1} = \frac{1}{2}\big(\xi^2-\xi \big), \hspace{5mm}
			p^{nn}_{*j} = \big(1-\xi^2\big), \hspace{5mm}
			p^{nn}_{*,\,j+1} =  \frac{1}{2}\big(\xi^2+\xi \big).
\end{equation}
Thus, the transaction probability from the grid point $x^{n-1}_i$ to grid point $x^n_j$ is obtained by a train rule that combines equations (\ref{4.4}) and (\ref{4.6}), as follows:
\begin{equation} \label{4.7}
		p^{n-1}_{ij} = \sum\limits_{*} p^{n-1}_{i*} \cdot p^{nn}_{*j}.
\end{equation}

The parameter $\mu^n$ in (\ref{state_map}) needs to be determined in order to map the grid point $x^n_i$ to an FX state $S^n_j = \exp(x^n_j + \mu^n)$ and calculate the volatility $\sigma^n_j = \sigma(t^n, S^n_j)$. This process is called curve calibration.
We begin the process at $t = t^0 = 0$, with only one state $S(x^0_0 + \mu^0)$ at the root grid $x^0_0$. Let this state be equal to the FX spot $F(t^0)$ published at $t^0$.
\begin{equation}
            S^0_0 = S(x^0_0 + \mu^0) = e^{x^0_0 + \mu^0} = F(t^0),
\end{equation}
we determine $\mu^0$.

According to the previous assumption that the local volatility function $\sigma = \sigma(t,S)$ is ready, we can calculate the volatility $\sigma(t^0,S^0_0)$ and transition probability $p^0_{0k}$. Denote the arriving probability at point $x^n_k$ by $q^n_{k}= p^{0n}_{0k}$. It can be seen that $q^1_{k} = p^0_{0k}$. Thus, the parameter $\mu^1$ can be determined by letting the expectation of the state FX at time $t^1$ be equal to the FX spot published at the same date $t^1$, namely,
\begin{equation}
            \sum\limits_{k = -I}^I p^0_{0k}\, S^1_k = \sum\limits_{k = -I}^I q^1_{k}\, S(x^1_k + \mu^1) = F(t^1).
\end{equation}

With $S^1_k = S(x^1_k+\mu^1)$, we have the volatility $\sigma(t^1,S^1_k)$ and the transition probability $p^1_{kj}$. Then, the probability of reaching the grid $x^2_j$ can be computed using the transition probability rule.
\begin{equation}
            q^2_j = \sum\limits_{k = -I}^I p^0_{0k}\, p^1_{kj} = \sum\limits_{k = -I}^I q^1_{k}\, p^1_{kj}
\end{equation}
Let the expectation of the FX state at date $t^2$ be equal to the FX spot published at the same date $t^2$, namely
\begin{equation}
            \sum\limits_{j = -I}^I q^2_{j}\, S^2_j = \sum\limits_{j = -I}^I q^2_{j}\, S(x^2_j + \mu^2) = F(t^2).
\end{equation}
We obtain $\mu^2$, as well as the grid state $S^2_j = S(x^2_j + \mu^2)$ at time $t^2$ for $j = -I, \cdots, I$.

We continue this process to calculate $\mu^3$, $\mu^4$, and so on until we reach the final time $t = T_e$. Then, we will finish the curve calibration and have the whole grid states $S^n_i = S(t^n, S_i)$, which can be used to calculate the value of American options.

The above calculation for grid states is called {\it forward propagation}. Once completed, a grid field $\{x^n_k\}$ is built. Each grid is mapped to an FX state $S^n_k = S(x^n_k+\mu^n)$, with which the option value can be calculated. For example, for a call option, if $x^{n+1}_j$ is a grid at the final time step of expiration date $t^{n+1} = T_e$, the option value is calculated by
\begin{equation} \label{opt_n+1}
                V^{n+1}_j = \max \big( S^{n+1}_j - K,\, 0).
\end{equation}
We are now moving back a step to another grid point $x^n_k$ at date $t^n$. Given that the transition probability $p^n_{kj}$ is known, the option values $V^{n+1}_j$ can be expressed as the following $\hat{V}^n_{k}$ from the viewpoint at $x^n_k$
\begin{equation} \label{roll_back}
                \hat{V}^n_{k} = \exp(-r^n \Delta t) \sum\limits_{j = -I}^I p^n_{kj}\, V^{n+1}_j,
\end{equation}
where, $r^n = r(t^n)$ is the interest rate of currency~2 (CNY) at date $t^n$, and $\hat{V}^n_{k}$ represents the option value if it is held on $x_k^n$ and exercised on a future date $t > t^n$.
If the option is exercised on $x^n_k$ at $t^n$, the immediate payoff is $\max(S^n_k-K,\, 0)$. To exercise the option now, this payoff must be greater than the future value $\hat{V}^n_{k}$. Otherwise, it is preferable to keep the option for the future. Thus, the option value at the grid point $x^n_k$ is the maximum of the current exercise value $\max(S^n_k-K,\, 0)$ and the future exercise value $\hat{V}^n_{k}$, namely
\begin{equation} \label{opt_n}
                V^{n}_k = \max \big( S^{n}_k - K,\, \hat{V}^n_{k}).
\end{equation}
It can be seen that equation (\ref{opt_n}) is equivalent to equation (\ref{opt_n+1}) except that we are now standing on the date $t^n$. We can use the process described in equation (\ref{roll_back}) by changing $n$ to $n-1$, and move the option value one step back to the date $t^{n-1}$. By repeating this process, we can propagate the option values of all grids back to the root grid $x^0_0$ and obtain the option PV, namely $V^0_0$. This calculation process is called {\it backward propagation}.

The forward propagation and backward propagation calculation method described above was proposed by Bill Krasker from Salomon Brothers in the 1990s. It is an explicit method and has first-order precision for $\Delta t$ and second-order precision for $\Delta x$. The Krasker method can be compared favorably to the Godunov method~[8] in fluid dynamics because they both have a clear physical background, fast computation speed, high numerical precision, and strong working stability. They are also easy to code. Additionally, the Krasker method can be applied to two- or three-dimensional grid computations.

\section{Monte-Carlo method for Asian options}

The value of an Asian option at time $t=T$ depends on the historical path of the underlying $S(t)$ in $t<T$. Therefore, a Monte Carlo method is used to price the Asian option. Unlike the grid method, where a symmetric grid setup benefits the computation of the transition probability, there is no need for symmetry in the Monte Carlo paths. Thus, the work is carried out based on the following model equation,
\begin{eqnarray} \label{mc_SDE}
          &&  {\rm d} x = -\frac{1}{2} \sigma^2(t,S) {\rm d}t +  \sigma (t,S) {\rm d} w,
          \nonumber\\[2mm]
          && S(t+{\rm d}t)  = S(t) e^{{\rm d}x + \mu},
\end{eqnarray}
where $\mu = \mu(t)$ is a parameter for curve calibration to the FX forward curve.

Similar to the grid method, if a correct volatility function $\sigma(t,S)$ is already known, the Monte-Carlo pricing process can be divided into three stages: path generation, curve calibration by forward propagation, and deal valuation. If the correct volatility function $\sigma(t,S)$ is not available, an approximate volatility function will be assumed, and the three-stage process will be used repeatedly to calibrate the volatility until the correct function is obtained. Once the correct function is obtained, the valuation work can begin. This section focuses on the three-stage process assuming the correct volatility function is available. The volatility calibration is addressed in Section 6.

In the first stage of path generation, we set up both the time step and path state.

We start with the time step, in which the date period $(0 \le t \le T_e)$ is divided into several small intervals. The work begins with collecting the expiration dates from market volatility instruments and the FX fixing dates from Asian option deals. These dates will form an initial time series. If an interval in the initial time series is too large (for example, greater than 3 days), we will add more dates in between.

Then, we move on to the path state. Suppose we prepare a number of $I$ paths in the pricing, all of which have a common time series $t^0, t^1, t^2,\cdots$. In the $i$-th path at date $t=t^n$, we take a random number $\xi^n_i$ from normal distribution for the Browning motion ${\rm d} w$ in (\ref{mc_SDE}),
\begin{equation} \label{mc_path}
            S^{n+1}_i =  S^{n}_i\exp\Big(-\frac{1}{2}\sigma^2(t^n,S^n_i) \Delta t + \sigma(t^n,S^n_i)
            \sqrt{\Delta t} \xi^n_i + \mu^n\Big).
\end{equation}
Thus we have a formula to generate the FX state  paths for  pricing.

The parameter $\mu^n$ in (\ref{mc_path}) needs to be determined to ensure that the state $S^{n+1}_j$ is correct and to calculate the volatility $\sigma^{n+1}_j = \sigma(t^{n+1}, S^{n+1}_j)$ at the next date $t^{n+1}$. This task is called curve calibration and must be done in conjunction with path generation. Starting from $n = 0$, all initial states $S^0_i$ of every path $i$ must be equal to the FX spot published on $t^0$,
\begin{equation}
            S^0_i = F(t^0), \hspace{1cm} (i = 1,2,\cdots,I).
\end{equation}
Suppose at date $t^n$ we already have the correct states $S^n_i$ and volatilities $\sigma^n_i = \sigma(t^n,S^n_i)$ for all paths $i = 1,2,\cdots,I$. We then project a test state at $t^{n+1}$ for each path $i$ by
\begin{eqnarray} \label{mc_path1}
            X^{n+1}_i &=&  S^{n}_i\exp\Big(-\frac{1}{2}\sigma^2(t^n,S^n_i) \Delta t + \sigma(t^n,S^n_i)
            \sqrt{\Delta t}\, \xi^n_i\Big), \nonumber \\[2mm]
            & &  (i = 1,2,\cdots,I).
\end{eqnarray}
The non-arbitrage condition requires that the expectation of $S^{n+1}_i$ be equal to the FX spot published on $t^{n+1}$, namely $F(t^{n+1})$. To satisfy this condition, let $\lambda = e^\mu$, and use the following parameter $\lambda^{n+1}$,
\begin{equation}
            \lambda^{n+1}  = F(t^{n+1})  \Big( \frac{1}{I} \sum\limits_{i=1}^I X^{n+1}_i \Big)^{-1},
\end{equation}
we obtain the non-arbitrage states at date $t^{n+1}$ as follows,
\begin{equation}
                S^{n+1}_i = \lambda^{n+1} \, X^{n+1}_i, \hspace{1cm} (i = 0, 1, \cdots, I).
\end{equation}
By repeating this work for $n+2$, $n+3$, and so on, we can complete curve calibration.

After calibrating the curve, we can begin valuing the Asian option deal. Along a given path $i$, we have FX states $S^n_i$ at every date $t^n$ where $n = 0, 1, 2, \cdots$. We can calculate the average FX fixing using (\ref{fixing}). The payoff of the option deal along this path $i$ can be obtained using  (\ref{payoff_asia_Arithmetic_Spot})-(\ref{payoff_asia_Geometric_Strike}). By taking an average of the payoff over all paths and discounting it from the payment date back to $t^0$, we obtain the deal present value (PV).

\section{Volatility calibration}

In order to perform the volatility calibration, we require an FX option pricer. The pricer can be constructed using either the grid method, as described in Section 4, or the Monte Carlo method, as described in Section 5.






\renewcommand\arraystretch{1.05}
\begin{table}[hbt]
\begin{center}
\caption{USDCNY Volatility Instruments on Valuation Date 20220926.}
\vspace{4mm}
\begin{small}
\begin{tabular}{|c|c|c|c|c|c|c|c|} \hline
\rule[0pt]{0pt}{5mm}
\tt	Vol  &\tt \ 10D\,Put\   &\tt \ 25D\,Put\  &\tt \ \	ATMF\ \	&\tt	25D\,Call	   &\tt	   10D\,Call	&\tt ExpDate &\tt PayDate	\\
\hline
\tt	1 Day	&\tt	-----	&\tt	-----	&\tt	6.4460	&\tt	-----	&\tt	-----	&\tt 20220927   &\tt 20220929 \\
\tt	1 Week	&\tt	-----	&\tt	4.2750	&\tt	4.2000	&\tt	4.2750	&\tt	-----	&\tt 20220930   &\tt 20221011 \\
\tt	2 Week	&\tt	3.4805	&\tt	3.3375	&\tt	3.2500	&\tt	3.3375	&\tt	3.7305	&\tt 20221010   &\tt 20221012 \\
\tt	3 Week	&\tt	3.8281	&\tt	3.6855	&\tt	3.5980	&\tt	3.6855	&\tt	4.0781	&\tt 20221017 	&\tt 20221019 \\
\tt 1 Month &\tt	4.3355	&\tt	4.1810	&\tt	4.0560	&\tt	4.1810	&\tt	4.5855	&\tt 20221026 	&\tt 20221028 \\
\tt	2 Month	&\tt	4.3397	&\tt	4.1644	&\tt	4.1130	&\tt	4.3144	&\tt	4.6897	&\tt 20221124 	&\tt 20221128 \\
\tt	3 Month	&\tt	4.6009	&\tt	4.4779	&\tt	4.4500	&\tt	4.7279	&\tt	5.1009	&\tt 20221226 	&\tt 20221228 \\
\tt	6 Month	&\tt	4.5013	&\tt	4.4060	&\tt	4.3500	&\tt	4.7560	&\tt	5.2013	&\tt 20230324 	&\tt 20230328 \\
\tt	9 Month	&\tt	4.3574	&\tt	4.3344	&\tt	4.3000	&\tt	4.7344	&\tt	5.5274	&\tt 20230626 	&\tt 20230628 \\
\tt	1 Year	&\tt	4.3114	&\tt	4.3238	&\tt	4.3000	&\tt	4.7488	&\tt	5.3114	&\tt 20230926 	&\tt 20230928 \\
\tt	18 Month &\tt	4.2841	&\tt	4.3260	&\tt	4.3000	&\tt	4.8010	&\tt	5.3341	&\tt 20240326 	&\tt 20240328 \\
\tt	2 Year	&\tt	4.2533	&\tt	4.2909	&\tt	4.3000	&\tt	4.8409	&\tt	5.3533	&\tt 20240925 	&\tt 20240927 \\
\tt	3 Year	&\tt	4.2536	&\tt	4.3420	&\tt	4.3000	&\tt	4.8920	&\tt	5.3536	&\tt 20250924 	&\tt 20250926 \\
\hline
\end{tabular}
\end{small}
\end{center}
\end{table}


Suppose there are $n$ free parameters in the local volatility surface $\sigma(t,S)$. In the example of Table~1, $n=198$, since there are $18$ maturities (day) and $11$ states (s), resulting in $18 \times 11 = 198$ free parameters. we represent these free parameters by a vector, namely
\begin{equation}
            {\bf x}  = ( x_1, x_2, \cdots, x_n)^T.
\end{equation}

Suppose there are $m$ market volatility instruments available. In this example, $m=59$, as shown in Table~2. For the {\tt 1 Day} expiration tenor, we select only the at-the-money forward {\tt ATMF} instrument, as the liquidity for out-of-the-money trades becomes very low within one business day. Similarly, for the {\tt 1 Week} tenor, we choose three out of five available instruments. The forward FX curve and CNY discount factor curve used for the calculation are shown in Table~3.






\renewcommand\arraystretch{1.05}
\begin{table}[hbt]
\begin{center}
\caption{USDCNY Forward FX Curve and CNY Discount Factor Curve on 20220926.}
\vspace{4mm}
\begin{small}
\begin{tabular}{|c|c|l||c|c|l|l|} \hline
\rule[0pt]{0pt}{5mm}
\tt  Days   &\tt Date    &\tt\ \  \ FX  &\tt 	Days    &\tt Date    &\tt\ \ \ Disc~Factor &\tt \ \  Zero~Rate  	\\
\hline
\tt	0   	&\tt 20220926	&\tt 7.091767	&\tt	0   	&\tt 20220926	&\tt 1 &\tt 0	  \\
\tt	1	    &\tt 20220927	&\tt 7.091267	&\tt	1   	&\tt 20220927	&\tt 0.99997030643729 &\tt 1.084573480510	  \\
\tt	2	    &\tt 20220928	&\tt 7.09055 	&\tt	92   	&\tt 20221227	&\tt 0.99594227324328 &\tt 1.614238917762	  \\
\tt	3	    &\tt 20220929	&\tt 7.0903125	&\tt	182  	&\tt 20230327	&\tt 0.99127367379894 &\tt 1.758944368042	   \\
\tt	15	    &\tt 20211011	&\tt 7.0888775	&\tt	274   	&\tt 20230627	&\tt 0.98596080974996 &\tt 1.884726241507	   \\
\tt	16	    &\tt 20221012	&\tt 7.08729	&\tt	366   	&\tt 20230927	&\tt 0.98005664202941 &\tt 2.010363044324	   \\
\tt	23	    &\tt 20221919	&\tt 7.085325	&\tt	732   	&\tt 20240927	&\tt 0.95528848515717 &\tt 2.282408600332	   \\
\tt	32	    &\tt 20221028	&\tt 7.08319	&\tt	1097   	&\tt 20250927	&\tt 0.92834798649655 &\tt 2.475463763546	   \\
\tt	63	    &\tt 20221128	&\tt 7.075425	&\tt	1826   	&\tt 20270926	&\tt 0.86914126651663 &\tt 2.805376130443	   \\
\tt	93      &\tt 20221228	&\tt 7.0664	&\tt	3653   	&\tt 20320926	&\tt 0.72952589429167 &\tt 3.153172511253	   \\
\tt	183	    &\tt 20230328	&\tt 7.04005	&\tt	      	&\tt 	&\tt\  &\tt 	   \\
\tt	275	    &\tt 20230628	&\tt 7.02575	&\tt	       	&\tt 	&\tt\  &\tt 	   \\
\tt	367	    &\tt 20230928	&\tt 7.007     &\tt	       	&\tt 	&\tt\  &\tt 	    \\
\tt	549	    &\tt 20240328	&\tt 7.00455	&\tt	    	&\tt 	&\tt\  &\tt 	   \\
\tt	732	    &\tt 20240927	&\tt 7.0013	&\tt	    	&\tt 	&\tt\  &\tt 	   \\
\tt	1096	&\tt 20250926	&\tt 7.03305	&\tt	       	&\tt 	&\tt\  &\tt 	   \\
\tt	1459	&\tt 20260924	&\tt 7.07605	&\tt	       	&\tt 	&\tt\  &\tt 	   \\
\tt	1824	&\tt 20270924	&\tt 7.01055	&\tt	     	&\tt 	&\tt\  &\tt 	   \\
\hline
\end{tabular}
\end{small}
\end{center}
\end{table}


Each instrument is a European call or put option with a given delta (or strike). Therefore, for each instrument, we can calculate a price using the Black-Scholes formula, given the forward FX curve and discount factor curve. Let $c_i, (i = 1,2,\cdots, m)$ be the price of the $i$-th instrument from the Black-Scholes formula (referred to as Market Price below).
For the same $i$-th instrument, if a local volatility surface ${\bf x}$ is given, we can also calculate a price using the grid method or Monte-Carlo method. We call this price the Model Price and denote it by $y_i = y_i({\bf x}),\, (i = 1,2,\cdots, m)$. We define a vector function by
\begin{equation}
            {\bf f}({\bf x}) = \Big(\frac{y_1({\bf x})}{c_1}-1,\,\frac{y_2({\bf x})}{c_2}-1,\, \cdots,\, \frac{y_m({\bf x})}{c_m}-1  \Big)^T.
\end{equation}
The objective of model calibration is to find a solution ${\bf x}=\hat{\bf x}$ for the equation
\begin{equation} \label{f(x)=0}
             {\bf f}({\bf x}) = {\bf 0}.
\end{equation}
The solution may not exist since (\ref{f(x)=0}) represents a non-linear function of ${\bf x}$. In general, the objective is to find a local minimum point ${\bf x}=\hat{\bf x}$ such that
\begin{equation} \label{mini}
            || {\bf f}(\hat{\bf x}) ||^2 =  {\bf f}^T(\hat{\bf x})\cdot {\bf f}(\hat{\bf x}) = {\rm minimum}.
\end{equation}

We use the Levenberg-Marquardt method to solve (\ref{f(x)=0}). The starting point of this method is the Newton method. Let ${\bf x}_k$ be an initial value of free parameters that we choose arbitrarily. We take a Taylor expansion for ${\bf f}({\bf x})$ in (\ref{f(x)=0}) and obtain
\begin{equation} \label{taylor_exp}
         {\bf f}({\bf x}) =  {\bf f}({\bf x}_k) + {\bf A}({\bf x}_k) ( {\bf x} - {\bf x}_k  ) = {\bf 0},
\end{equation}
where ${\bf A}({\bf x}_k)$ is the Jacobian matrix defined as
\begin{equation}
            {\bf A}({\bf x}_k) = \frac{ \partial {\bf f}({\bf x}_k) } { \partial {\bf x}^T },
\end{equation}
and the partial derivatives in the formulas are calculated numerically using finite difference perturbation. From (\ref{taylor_exp}) we obtain an iteration formula for calculating the root
\begin{equation} \label{newton_soln}
            {\bf x}_{k+1} = {\bf x}_k - \big( {\bf A}^T {\bf A} \big)^{-1} {\bf A}^T {\bf f} ({\bf x}_k).
\end{equation}
If the iteration converges, the error will become smaller and smaller, namely,
\begin{equation}
				{\bf f}^T({\bf x}_{k+1}) \cdot {\bf f}({\bf x}_{k+1})  \le {\bf f}^T({\bf x}_{k}) \cdot {\bf f}({\bf x}_{k}).
\end{equation}
However, the matrix $\big( {\bf A}^T {\bf A} \big)$ may not be invertible, which can cause the iteration to fail. Even if the matrix is invertible, the moving step $({\bf x}_{k+1} - {\bf x}_k)$ obtained by (\ref{newton_soln}) may be too large to control the iteration process. Therefore, we can use the Levenberg-Marquardt method to perform the task. Let ${\bf w} = {\bf x} - {\bf x}_k$, and ${\bf f}_k ={\bf f}({\bf x}_k)$. We can rewrite the (\ref{taylor_exp}) into another form,
\begin{equation}
                {\bf A} {\bf w} = - {\bf f}_k.
\end{equation}
Multiplying this equation by ${\bf A}^T$, and adding a diagonal element $\alpha$, yields
\begin{equation}
                \big({\bf A}^T{\bf A} + \alpha {\bf I} \big) {\bf w} = - {\bf A}^T{\bf f}_k + \alpha {\bf w}.
\end{equation}
In the Levenberg-Marquardt method, we first select a sufficiently large positive number $\alpha$ to ensure that the matrix $\big({\bf A}^T{\bf A} + \alpha {\bf I} \big)$ is positive-definite. We then solve the equation
\begin{equation}
                \big({\bf A}^T{\bf A} + \alpha {\bf I} \big) {\bf w} = - {\bf A}^T{\bf f}_k
\end{equation}
for a solution ${\bf w}$. At this time, we  perform two checks. The first check is to verify whether ${\bf w}$ is still in the gradient direction of the original equation,  namely,
\begin{equation}
    {\bf w}^T \cdot \big(- {\bf A}^T{\bf f}_k + \alpha {\bf w}\big) > 0.
\end{equation}
The second check is whether ${\bf w}$ reduces the error, namely,
\begin{equation}
				{\bf f}^T({\bf x}_{k} + {\bf w}) \cdot {\bf f}({\bf x}_{k} + {\bf w})  \le {\bf f}^T({\bf x}_{k}) \cdot {\bf f}({\bf x}_{k}).
\end{equation}
If both conditions are met, ${\bf w}$ is valid. We will use ${\bf x}_{k+1} = {\bf x}_{k} + {\bf w}$ as the new initial condition for the next iteration, and decrease $\alpha$ by half. If at least one condition fails, we double $\alpha$ and test again until we find a valid ${\bf w}$. After several cycles of iterations, if $||{\bf f}({\bf x}_{k+1}) ||$ decreases to 0, we get the exact solution $\hat{\bf x} = {\bf x}_{k+1}$. Otherwise, if $||{\bf w} || = || {\bf x} - {\bf x}_k||$ becomes very small, we also get an optimizing solution ${\bf x}_{k+1}$. Thus, the volatility calibration is finished.





\renewcommand\arraystretch{1.05}
\begin{table}
\begin{center}
\caption{Result of Volatility Calibration by LM/Grid Method,  AvgError = 0.0003.}
\vspace{4mm}
\begin{small}
\begin{tabular}{|c|c|c|c|c|c|} \hline
\rule[0pt]{0pt}{5mm}
\tt	\ \ Instr\ \ \  &\tt \	MarketVol	\     &\tt	\ \ Strike \ \   &\tt  	MarketPrice 	&\tt\,	ModelPrice	   &\tt	\ \   Error	\ \ \\
\hline
\tt 1D\_ATM	&\tt	6.4460	&\tt	7.09031	&\tt\	\ 95.3935	&\tt\	\ 95.3925	&\tt	 -0.0000	\\
\tt	1W\_25P	&\tt	4.2750	&\tt	7.06759	&\tt\	\ 47.3778	&\tt\	\ 47.3772	&\tt	 -0.0000	\\
\tt	1W\_ATM	&\tt	4.2000	&\tt	7.08888	&\tt\	124.2192	&\tt\    124.2183	&\tt	 -0.0000	\\
\tt	1W\_25C	&\tt	4.2750	&\tt	7.11037	&\tt\	\ 47.1660	&\tt\	\ 47.1644	&\tt	 -0.0000	\\
\hline
\tt	2W\_10P	&\tt	3.4805	&\tt	7.02583	&\tt\	\ 22.9127	&\tt\	\ 22.9131	&\tt	\ 0.0000	\\
\tt	2W\_25P	&\tt	3.3375	&\tt	7.05627	&\tt\	\ 69.2511	&\tt\	\ 69.2514	&\tt	\ 0.0000	\\
\tt	2W\_ATM	&\tt	3.2500	&\tt	7.08729	&\tt\	 179.7794	&\tt\	 179.7761   &\tt	\ 0.0000	\\
\tt	2W\_25C	&\tt	3.3775	&\tt	7.11875	&\tt\	\ 68.7994	&\tt\	\ 68.7923	&\tt	 -0.0001	\\
\tt	2W\_10C	&\tt	3.7305	&\tt	7.15413	&\tt\	\ 24.4149	&\tt\	\ 24.4158	&\tt	\ 0.0000	\\
\hline
\tt	3W\_10P	&\tt	3.8281	&\tt	7.00276	&\tt\	\ 30.8771	&\tt\	\ 30.8800	&\tt	\ 0.0001	\\
\tt	3W\_25P	&\tt	3.6855	&\tt	7.04349	&\tt\	\ 93.7119	&\tt\   \ 93.7118	&\tt	 -0.0000	\\
\tt	3W\_ATM	&\tt	3.5980	&\tt	7.08533	&\tt\	 243.6170	&\tt\    243.6165	&\tt	 -0.0000	\\
\tt	3W\_25C	&\tt	3.6855	&\tt	7.12796	&\tt\	\ 92.8860	&\tt\     92.8874	&\tt	\ 0.0000	\\
\tt	3W\_10C	&\tt	4.0781	&\tt	7.17502	&\tt\	\ 32.6356	&\tt\	\ 32.6278	&\tt	 -0.0002	\\
\hline
\tt	1M\_10P	&\tt	4.3355	&\tt	6.97183	&\tt\	\ 41.8243	&\tt\	\ 41.8281	&\tt	\ 0.0001	\\
\tt	1M\_25P	&\tt	4.1810	&\tt	7.02688	&\tt\	 127.1780	&\tt\    126.9918	&\tt	 -0.0015	\\
\tt	1M\_ATM	&\tt	4.0560	&\tt	7.08319	&\tt\	 328.0121	&\tt\    328.0098	&\tt	 -0.0000	\\
\tt	1M\_25C	&\tt	4.1810	&\tt	7.14118	&\tt\	 125.6607	&\tt\    125.6466	&\tt	 -0.0001	\\
\tt	1M\_10C	&\tt	4.5855	&\tt	7.20412	&\tt\	\ 43.7688	&\tt\	\ 43.8282	&\tt	\ 0.0014	\\
\hline
\tt	2M\_10P	&\tt	4.3397	&\tt	6.92008	&\tt\	\ 58.6873	&\tt\	\ 58.6866	&\tt	 -0.0000	\\
\tt	2M\_25P	&\tt	4.1644	&\tt	6.99698	&\tt\	 177.6282	&\tt\    177.6174	&\tt	 -0.0001	\\
\tt	2M\_ATM	&\tt	4.1130	&\tt	7.07453	&\tt\	 465.3075	&\tt\    465.3436	&\tt	\ 0.0001	\\
\tt	2M\_25C	&\tt	4.3144	&\tt	7.15974	&\tt\	 180.9127	&\tt\    180.9211	&\tt	\ 0.0000	\\
\tt	2M\_10C	&\tt	4.6897	&\tt	7.24970	&\tt\	\ 62.4719	&\tt\	\ 62.4791	&\tt	\ 0.0001	\\
\hline
\tt	3M\_10P	&\tt	4.6009	&\tt	6.86372	&\tt\	\ 77.2500	&\tt\	\ 77.2685	&\tt	\ 0.0002	\\
\tt	3M\_25P	&\tt	4.4779	&\tt	6.96241	&\tt\	 237.2618	&\tt\    237.2635	&\tt	\ 0.0000	\\
\tt	3M\_ATM	&\tt	4.4500	&\tt	7.06640	&\tt\	 623.5861	&\tt\    623.5448	&\tt	 -0.0001	\\
\tt	3M\_25C	&\tt	4.7279 	&\tt    7.18178	&\tt\    244.8102	&\tt\    244.8232   &\tt    \ 0.0001	\\
\tt	3M\_10C	&\tt	5.1009  &\tt    7.30314 &\tt\   \ 83.9403	&\tt\	\ 83.9400	&\tt	 -0.0000	\\
\hline
\tt	6M\_10P	&\tt	4.5013	&\tt	6.76477	&\tt	\ 105.4837	&\tt	\ 105.4759	&\tt	 -0.0001	\\
\tt	6M\_25P	&\tt	4.4060	&\tt	6.89838	&\tt	\ 326.0658	&\tt	\ 326.0686	&\tt	\ 0.0000	\\
\tt	6M\_ATM	&\tt	4.3500	&\tt	7.04005	&\tt	\ 847.7307	&\tt	\ 847.8345	&\tt	\ 0.0001	\\
\tt	6M\_25C	&\tt	4.7560	&\tt	7.20393	&\tt	\ 340.8460	&\tt	\ 340.8558	&\tt	\ 0.0000	\\
\tt	6M\_10C	&\tt	5.2013	&\tt	7.38124	&\tt	\ 118.4995	&\tt	\ 118.5032	&\tt	\ 0.0000	\\
\hline
\tt	9M\_10P	&\tt	4.3574	&\tt	6.69937	&\tt	\ 125.4967	&\tt	\ 125.4916	&\tt     -0.0000	\\
\tt	9M\_25P	&\tt	4.3344	&\tt	6.85521	&\tt	\ 394.5328	&\tt	\ 394.4917	&\tt	 -0.0001	\\
\tt	9M\_ATM	&\tt	4.3000	&\tt	7.02575	&\tt	 1027.2206	&\tt	 1027.0325	&\tt	 -0.0002	\\
\tt	9M\_25C	&\tt	4.7344	&\tt	7.22847	&\tt	\ 414.3587	&\tt	\ 414.3456	&\tt	 -0.0000	\\
\tt	9M\_10C	&\tt	5.2574	&\tt	7.45485	&\tt	\ 146.2854	&\tt	\ 146.2823	&\tt	 -0.0000	\\
\hline
\end{tabular}
\end{small}
\end{center}
\end{table}

\setcounter{table}{3}
\begin{table}[t]
\begin{center}
\caption{(Continued)}
\vspace{4mm}
\begin{small}
\begin{tabular}{|c|c|c|c|c|c|}
\hline \rule[0pt]{0pt}{5mm}
\tt	\ \ Instr\ \ \  &\tt \	MarketVol	\     &\tt	\ \ Strike \ \   &\tt  	MarketPrice 	&\tt\,	ModelPrice	   &\tt	\ \   Error	\ \ \\
\hline
\tt	1Y\_10P	&\tt	4.3114	&\tt	6.63663	&\tt	\ 142.6636	&\tt	\ 142.7026	&\tt    \ 0.0003	\\
\tt	1Y\_25P	&\tt	4.3238	&\tt	6.81203	&\tt	\ 452.4687	&\tt	\ 452.4639	&\tt	 -0.0000	\\
\tt	1Y\_ATM	&\tt	4.3000	&\tt	7.00700	&\tt     1177.4686	&\tt	 1177.2818	&\tt	 -0.0002	\\
\tt	1Y\_25C	&\tt	4.7488	&\tt	7.24315	&\tt	\ 474.8915	&\tt	\ 474.9302	&\tt	\ 0.0001	\\
\tt	1Y\_10C	&\tt	5.3114	&\tt	7.51097	&\tt	\ 168.8805	&\tt	\ 168.8806	&\tt	\ 0.0000	\\
\hline
\tt	18M\_10P&\tt	4.2841	&\tt	6.55839	&\tt	\ 171.9610	&\tt	\ 171.9223	&\tt     -0.0002	\\
\tt	18M\_25P&\tt	4.3260	&\tt	6.76832	&\tt	\ 549.7037	&\tt	\ 549.7591	&\tt	 -0.0001	\\
\tt	18M\_ATM&\tt	4.3000	&\tt	7.00455	&\tt     1422.6565	&\tt	 1422.3343	&\tt	 -0.0002	\\
\tt	18M\_25C&\tt	4.8010	&\tt	7.30029	&\tt	\ 576.9045	&\tt	\ 577.0090	&\tt	\ 0.0002	\\
\tt	18M\_10C&\tt	5.3341	&\tt	7.63197	&\tt	\ 203.9122	&\tt	\ 203.9107	&\tt	 -0.0000	\\
\hline
\tt	2Y\_10P	&\tt	4.2533	&\tt	6.49377	&\tt	\ 195.2634	&\tt	\ 195.2728	&\tt    \ 0.0000	\\
\tt	2Y\_25P	&\tt	4.2909	&\tt	6.73300	&\tt	\ 624.0373	&\tt	\ 624.0487	&\tt	\ 0.0000	\\
\tt	2Y\_ATM	&\tt	4.3000	&\tt	7.00130	&\tt     1621.7772	&\tt	 1621.8055	&\tt	\ 0.0000	\\
\tt	2Y\_25C	&\tt	4.8409	&\tt	7.34925	&\tt	\ 660.0036	&\tt	\ 660.0085	&\tt	\ 0.0000	\\
\tt	2Y\_10C	&\tt	5.3533	&\tt	7.73650	&\tt	\ 232.3157	&\tt	\ 232.3095	&\tt	 -0.0000	\\
\hline
\tt	3Y\_10P	&\tt	4.2536	&\tt	6.41724	&\tt	\ 234.7247	&\tt	\ 234.7244	&\tt     -0.0000	\\
\tt	3Y\_25P	&\tt	4.3420	&\tt	6.70437	&\tt	\ 760.3419	&\tt	\ 760.3198	&\tt	\ 0.0000	\\
\tt	3Y\_ATM	&\tt	4.3000	&\tt	7.03305	&\tt     1938.1189	&\tt	 1938.0847	&\tt	 -0.0000	\\
\tt	3Y\_25C	&\tt	4.8920	&\tt	7.47310	&\tt	\ 790.8628	&\tt	\ 790.8363	&\tt	 -0.0000	\\
\tt	3Y\_10C	&\tt	5.3536	&\tt	7.95382	&\tt	\ 275.7694	&\tt	\ 275.7730	&\tt	\ 0.0000	\\
\hline
\end{tabular}
\end{small}
\end{center}
\end{table}


Table 4 shows a result of the volatility calibration, where the error of each instrument is defined as
\begin{equation}
            {\rm Error}_i = f_i = \frac{y_i({\bf x})}{c_i} - 1, \hspace{1cm} (i = 1,2, \cdots,m),
\end{equation}
and the average calibration error is
\begin{equation}
        {\rm AvgError} = \sqrt{ \frac{1}{m} {\bf f}^T({\bf x}) \cdot {\bf f}({\bf x})}
            = \sqrt{\frac{1}{m} \sum\limits_{i=1}^m \Big( \frac{y_i({\bf x})}{c_i}-1 \Big)^2 } = 0.0003.
\end{equation}
The above volatility calibration was performed using the grid method. We found that the accuracy of the calibration depends mainly on the smoothness of the market instrument volatility surface (Table~2). For this calibration data, the result is equally good if the Monte-Carlo method is used. However, for unrealistic/arbitrageable market volatilities, the calibration result may be worse. Based on our research, the market volatility surface must meet at least the following two conditions for a good calibration:

(C1): The variances of ATM volatilities increase with expiration tenors. For example, from Table 2, the 1D ATM volatility is 6.446\%, the 1W ATM volatility is 4.2\% (since ExpDate = 20220930, it has only 4 days), the 2W ATM volatility is 3.25\% (14 days), and so on. Then this condition  asks
\begin{equation}
             6.446^2 \times 1 \le 4.2^2 \times 4 \le 3.25^2 \times 14 \le \cdots.
\end{equation}

(C2): For every expiration tenor, the BF25 volatility is greater than zero and less than the BF10 volatility. For example, for the 1Y tenor in Table~2,
\begin{eqnarray}
         {\rm BF25} &=& \frac{1}{2} \big( 4.3238 + 4.7488) - 4.3 = 0.2363, \nonumber\\[2mm]
            {\rm BF10} &=& \frac{1}{2} \big( 4.3114 + 5.3114) - 4.3 = 0.5114.
\end{eqnarray}
Thus, we have $0 \le {\rm BF25} \le {\rm BF10}$. This condition must also be held for other tenors.

\section{Pricing precision}

Once the volatility calibration is properly done, we can use the local volatility function $\sigma(t, S)$ to construct a pricer and evaluate option deal prices. Although the pricer, whether built using grid or Monte-Carlo methods, can deliver accurate prices for European options, calculation errors are unavoidable when pricing American or Asian options. The following are two examples that demonstrate calculation errors caused by the number of grids (or number of Monte-Carlo paths) used in pricing.





\renewcommand\arraystretch{1.05}
\begin{table}[hbt]
\begin{center}
\caption{The Effect of Grid Number on American Options. Pricing Date:\,20220926.}
\vspace{4mm}
\begin{small}
\begin{tabular}{|c||c|c|c|c|} \hline
\rule[0pt]{0pt}{5mm}
\tt \	Deal \   &\tt 1Y\_ATM    &\tt 1Y\_10C  &\tt 6M\_ATM	&\tt	1M\_ATM	 	\\
\hline
\tt	I=50	&\tt 1290.4302	&\tt 173.9357	&\tt 920.4592	&\tt 331.7334	 \\
\tt	I=100	&\tt 1290.1247	&\tt 173.9330	&\tt 920.2302	&\tt 331.7870	 \\
\tt	I=200	&\tt 1290.2568	&\tt 173.9315	&\tt 920.1014	&\tt 331.8036	 \\ \hline
\tt	Strike	&\tt 7.00700	&\tt 7.51097	&\tt 7.04005	&\tt 7.08319	 \\
\tt	~ExpDate~ 	&\tt 20230926	&\tt 20030926	&\tt 20030324	&\tt 20021026	 \\
\tt	PayDate	&\tt 20230928	&\tt 20030928	&\tt 20030328	&\tt 20021028	 \\
\tt	EuroPrc	&\tt 1177.4686	&\tt 168.8805	&\tt 847.7307	&\tt 328.0121	 \\
\hline
\end{tabular}
\end{small}
\end{center}
\end{table}


Table~5 shows the values of four American option deals calculated using the grid method with varying numbers of half-grids. The {\tt 1Y\_ATM} deal is an extension of the 1-year ATM European option instrument in Table~4, and the same applies to the other three deals. The grid method was used to price these four American options, with half-grid numbers $I$ of 50, 100, and 200. The pricing values are listed in the table. The results demonstrate that the grid method is highly stable, with calculation results for a given deal differing by no more than 1 basis point across different values of $I$.





\renewcommand\arraystretch{1.05}
\begin{table}[hbt]
\begin{center}
\caption{The Effect of Path Number on Asian Options. Pricing Date:\,20220926.}
\vspace{4mm}
\begin{small}
\begin{tabular}{|c||c|c|c|c|} \hline
\rule[0pt]{0pt}{5mm}
\tt \	Deal \   &\tt 1Y\_ATM    &\tt 1Y\_10C  &\tt 6M\_ATM	&\tt	1M\_ATM	 	\\
\hline
\tt	I=5000	&\tt 883.2436	&\tt 11.8253	&\tt 608.9964	&\tt 186.5733	 \\
\tt	I=10000	&\tt 881.1290	&\tt 12.0545	&\tt 612.8797	&\tt 186.9654	 \\
\tt	I=15000	&\tt 878.8648	&\tt 11.9601	&\tt 612.0276	&\tt 186.9369	 \\
\tt	I=20000	&\tt 881.0790	&\tt 12.6660	&\tt 611.2084	&\tt 187.3059	 \\
 \hline
\tt	Strike	&\tt 7.00700	&\tt 7.51097	&\tt 7.04005	&\tt 7.08319	 \\
\tt	~LastFix~ 	&\tt 20230926	&\tt 20030926	&\tt 20030324	&\tt 20021026	 \\
\tt	PayDate	&\tt 20230928	&\tt 20030928	&\tt 20030328	&\tt 20021028	 \\
\tt	EuroPrc	&\tt 1177.4686	&\tt 168.8805	&\tt 847.7307	&\tt 328.0121	 \\
\hline
\end{tabular}
\end{small}
\end{center}
\end{table}


Table 6 shows the values of four Asian option deals calculated using the Monte-Carlo method with different numbers of paths, $I$. Each deal was constructed from the extension of the European option instrument of the same name in Table 4. For all four deals, the FX fixing dates were set to Fridays, or the following CFETS business day if Friday is a holiday. The numbers of paths $I$ were taken as 5000, 10000, 15000, and 20000, respectively.
From the table, we see that for short-term deals such as {\tt 1M\_ATM} and out-of-the-money deals like {\tt 1Y\_10C}, the prices are stable within 1 basis point. Even for the long-term deal like {\tt 1Y\_ATM}, there is only about a 2 basis point variation in the prices, which basically satisfies the business requirements.
Moreover, to further improve pricing precision, we can increase the number of paths $I$, increase the number of local volatility grids (currently $18\times 11$), or increase the number of time steps.

\section{Local-stochastic volatility model}

The local volatility model (LV), proposed in this paper, can be easily extended to the local-stochastic volatility model (LSV). For example, if the Heston model is used to formulate the stochastic volatility feature, the underlying FX rate $S$ and its variance rate $V$ are driven by two correlated Brownian motions, namely, $w_1$ and $w_2$,
\begin{eqnarray} \label{lsv}
    \frac{{\rm d} S}{S} &=&  \sigma(t,S) \sqrt{V} {\rm d} w_1,
    \nonumber\\[2mm]
    {\rm d} V &= &\kappa(\theta - V){\rm d} t + \varepsilon \sqrt{V}  {\rm d} w_2,
    \nonumber\\[2mm]
   && {\rm d} w_1 \cdot {\rm d} w_2 = \rho {\rm d} t.
\end{eqnarray}
where,   $\kappa$, $\theta$, $\varepsilon$, $\rho$, $V_0 = V(0)$ (not shown in the formulas) are the five free parameters of the Heston model.

In the history of model development, a major objective of LSV has been to have more free model parameters in order to calibrate the strong market volatility skew, which is shown in Table 2, for example. There have been many achievements in LSV research, as seen in references [1-7].
Based on these references and our work in this paper, we would like to make two comments.
Firstly, the LSV is a two-factor model which requires a two-dimensional grid or Monte-Carlo numerical method. The coding complexity and computation time will increase to a multiple level compared to the current LV model. Secondly, no matter how the local volatility function $\sigma(t,S)$ is selected, volatility calibrations always yield poor results if there are not enough free parameters to support it. For instance, [7] presents a scenario in which 30 free parameters are chosen in $\sigma(t,S)$ to calibrate 30 volatility instruments, and the ${\rm AvgError}$ is 0.0177. Clearly, this result falls short of meeting the requirements of trading businesses.

The fully parameterized one-factor local volatility model proposed in this paper can be accurately and efficiently calibrated to the volatility skew, which meets the requirements of our trading desk. Therefore, there is no need to extend this model to the two-dimensional LSV model.

\vspace{4mm}
{\bf Acknowledgement}: We would like to express our sincere thanks to our colleagues CHEN Xuejun, WEN Yang, and LI Yipeng from the FX trading division of CCB Financial Market Department for their significant contributions in developing, validating, and working with the LV model in this paper. We could not have achieved success without their support.

\section*{\large Reference}

\begin{enumerate}

\item[1.] C. Homescu: Local stochastic volatility models: calibration and pricing. \\ http://www.researchgate.net/publication/272247504, (2014).

\item[2.] E. Derman and I. Kani: Stochastic implied trees: arbitrage pricing with stochastic term and strike structure of volatility. {\it Goldman Sachs Quantitative Strategies Research Notes},
http://www.researchgate.net/publication/240264588, (1997).

\item[3.] S.L. Heston: A closed-form solution for options with stochastic volatility with
applications to bond and currency options. {\it The Review of Financial Studies},  6  327-343 (1993).

\item[4.] L. Andersen: Simple and efficient simulation of the Heston stochastic volatility model. Journal of Computational Finance, 11(3) 1-42  (2008).

\item[5.] A. van der Stoep, L.A. Grzelak, C.W. Oosterlee:
The Heston Stochastic-Local Volatility Model: Efficient Monte Carlo Simulation,
{\it International Journal of Theoretical and Applied Finance}, 17(7)  (2014).

\item[6.] U.P. Wystup: The Tremor Model The Tremor Stochastic-Local-Volatility
Model Independent Validation by MathFinance. http://researchgate.net/publication/265408143.

\item[7.] X. Lin: The Pricing Models for CNY Structured Deposit
Trades (in Chinese), (2019).\\ http://www.researchgate.net/publication/336275297.

\item[8.] X. Lin and J. Ballmann: A Riemann solver and a second-order Godunov method for elastic-plastic wave propagation in solids, {\it International Journal of Impact Engineering},  13(3) 463-478
(1993). http://www.researchgate.net/publication/245148831.

\end{enumerate}

\end{document}